\documentclass[aps,prd,amsmath,amssymb,preprintnumbers,nofootinbib,a4paper,11pt]{revtex4}
\pdfoutput=1
\usepackage{amsthm}
\usepackage{graphicx}
	\graphicspath{{./NewFig/}}
\usepackage{url}
\usepackage[bookmarks, pagebackref=false]{hyperref}
\usepackage{color}
	\definecolor{rossoCP3}{cmyk}{0,.88,.77,.40}
		\definecolor{graa}{rgb}{0.8,0.8,0.8}
		\definecolor{blaa}{rgb}{0.2,0.2,0.6}
		\hypersetup{
			colorlinks, 
			bookmarksopen, 
			bookmarksnumbered,
			citecolor=blaa, 		
			linkcolor=rossoCP3,	
			urlcolor=rossoCP3,			
			}
\usepackage{dcolumn}
\usepackage{bm}
\usepackage{bbm}
\usepackage{pxfonts}

\usepackage{epsfig}
\usepackage{placeins}

\usepackage[ margin=5pt, font=small,labelfont=bf, justification=raggedright]{caption}
\usepackage{slashed}
\usepackage{subfig}

\newcommand{\beq}{\begin{eqnarray}}
\newcommand{\eeq}{\end{eqnarray}}

\newcommand{\bmp}{\noindent\begin{minipage}{16cm}}
\newcommand{\emp}{\end{minipage}\vskip 7mm} 

\def\lsim{\mathrel{\rlap{\lower4pt\hbox{\hskip1pt$\sim$}}
    \raise1pt\hbox{$<$}}}                
\def\gsim{\mathrel{\rlap{\lower4pt\hbox{\hskip1pt$\sim$}}
    \raise1pt\hbox{$>$}}}                

\begin{document}
\title{\Large  \color{rossoCP3}  Large N Scalars: From Glueballs  to Dynamical Higgs Models} 
 \author{Francesco Sannino
 }
 \email{sannino@cp3.dias.sdu.dk} 
  \affiliation{
{  CP}$^{ 3}${-Origins} \& the Danish IAS,
University of Southern Denmark, Campusvej 55, DK-5230 Odense M, Denmark.
}

\begin{abstract}
We construct effective Lagrangians, and corresponding counting schemes, valid to describe the dynamics of the lowest lying large N stable massive composite state emerging in strongly coupled theories. The large N counting rules can now be employed when computing quantum corrections via an effective Lagrangian description. The framework allows for systematic investigations of  composite dynamics of non-Goldstone nature. Relevant examples are 
 the lightest glueball states emerging in any Yang-Mills theory. We further apply the effective approach and associated counting scheme to  composite models at the electroweak scale.  To illustrate the formalism we consider the possibility that the Higgs emerges as: the lightest glueball of a new composite theory; the large N scalar meson in models of dynamical electroweak symmetry breaking;  the large N pseudodilaton useful also for models of near-conformal dynamics. For each of these realisations we determine the leading N corrections to the electroweak precision parameters. The results nicely elucidate the underlying large N dynamics and can be used to confront first principle lattice results featuring composite scalars with a systematic effective approach.     \vskip .1cm
{\footnotesize  \it Preprint: CP$^3$-Origins-2015-035 DNRF90\& DIAS-2015-35}
 \end{abstract}

\maketitle

\section{Introduction}

Strong dynamics continues to pose a formidable challenge.  Several analytical and numerical ingenious techniques have been invented, exploited and are routinely used to elucidate some of its physical properties. The large number of underlying colours limit is a time-honoured example  \cite{'tHooft:1973jz,Witten:1979kh,Witten:1980sp}. It has been extensively used in Quantum Chromodynamics (QCD) and string theory, and it constitutes the backbone of the gauge-gravity duality program. We will further highlight here the power of the large $N$ expansion by introducing a four-dimensional calculable framework permitting to investigate the dynamics of the lightest stable non-Goldstone  large $N$ composite state.

 t' Hooft and Witten demonstrated that Yang-Mills theories at large number of colours admit an effective description in terms of an infinite number of non-interacting absolutely stable hadronic states of arbitrary spin \cite{'tHooft:1973jz,Witten:1979kh,Witten:1980sp}. By capitalising on this central result we focus on the physics of the lightest massive scalar state that is known to play an important role in QCD \cite{Sannino:1995ik,Harada:1995dc,Harada:1996wr,Harada:2003em,Black:2000qq,Caprini:2005zr,GarciaMartin:2011jx,Parganlija:2012fy,Pelaez:2015zoa,Ghalenovi:2015mha} and in various models of dynamical electroweak symmetry breaking as summarised in \cite{Sannino:2009za}.  The framework can also be used to consistently determine  quantum corrections to compare with first principle lattice simulations of composite dynamics featuring scalars \cite{Bursa:2011ru,Fodor:2015vwa,Kuti:2014epa,Fodor:2012ty}.

The paper is organised as follows. In Section~\ref{LGB} we introduce the effective theory for the lightest massive glueball scalar state emerging within a pure Yang-Mills theory, and provide the associated counting scheme. Here we discuss the intriguing interplay between momentum and large $N$ expansions. The framework goes beyond the glueball example and lays the foundation of the subsequent analyses. We then extend the framework to several models of composite electroweak dynamics in Section~\ref{DESB} where the Higgs is identified with the lightest composite state. In Section~\ref{ST} we determine the large $N$ dependence, for each model, of the electroweak precision parameters \cite{Peskin:1990zt} stemming  from different dynamical Higgs realisations. We summarise our results in Section~\ref{conclusions}.

\section{Large N Effective Theory For the Lightest Glueball State}
 \label{LGB}
 
Consider the lightest scalar state stemming from an $SU(N)$ pure Yang-Mills theory, which is also expected to be the lowest lying state of the full theory. At infinite number of colours the effective Lagrangian for this state is simply the one of a free scalar field \cite{'tHooft:1973jz,Witten:1979kh,Witten:1980sp}
\begin{eqnarray} 
L_{GB} = \frac{1}{2} \partial_{\mu} h \partial^{\mu} h - \frac{m^2_h}{2} h^2 + {\cal O}(N^{-1})\ . 
\end{eqnarray}
It is possible to go beyond the free-field limit by first defining with $\Lambda_H$  the intrinsic composite scale of the theory that permits to expand the effective Lagrangian both in $1/N$ and $\partial^2/\Lambda_H^2$ as follows
\begin{equation}
L_{GB} = \frac{1}{2} \partial_{\mu} h \partial^{\mu} h 
 -  \frac{m^2_h}{2} h^2 \sum_{q=0,p=0}^\infty V_{q,p}\left(\frac{\partial^2}{\Lambda_H^2}\right)^q\left(\frac{1}{N}\frac{h}{ \Lambda_H}\right)^p \ . 
\end{equation}
Here $V_{q,p}$ are dimensionless coefficients of order unity with $V_{0,0}=1$.  The expansion in $1/N$ takes care of the large $N$ suppression of higher point correlators, while the higher derivative terms take into account integrating out heavier states.  Since the heavier states couple via $1/N$ suppressed interactions we must also have $V_{q,0} = 0$ with $q\geq 1$. 

To leading order in the double expansion we have: 
\begin{equation}
L_{GB} = \frac{1}{2} \partial_{\mu} h \partial^{\mu} h 
-  \frac{m^2_h}{2} h^2 \left[ 1 + \frac{V_{0,1}}{N}\frac{h}{ \Lambda_H}  \right] \ . 
\end{equation}
This shows that the trilinear coupling of the scalar is naturally suppressed in this limit. Expanding a little further we have: 
\begin{eqnarray}
L_{GB} = \frac{1}{2} \partial_{\mu} h \partial^{\mu} h 
 -  \frac{m^2_h}{2} h^2 \left[ 1 +  \frac{V_{0,1}}{N}\frac{h}{ \Lambda_H} +  \frac{V_{1,1}}{N} 
\frac{\partial^2}{\Lambda_H^2}\frac{h}{ \Lambda_H}  +\frac{V_{0,2}}{N^2}\frac{h^2}{ \Lambda_H^2}  \right] \ . \nonumber \\ 
\end{eqnarray}
We have ordered the terms in such a way that the $1/N$ order counts as $\partial^2/\Lambda^2_H$, however, we can imagine several different situations. For example  we can work in the very low momentum region. In this limit we can order $\partial^2/\Lambda_H^2 \sim 1/N^2$ and therefore to the next leading order we drop the derivative terms and obtain
\begin{eqnarray}
L_{GB} = \frac{1}{2} \partial_{\mu} h \partial^{\mu} h 
-  \frac{m^2_h}{2} h^2 \left[ 1 +  \frac{V_{0,1}}{N}\frac{h}{ \Lambda_H}   +\frac{V_{0,2}}{N^2}\frac{h^2}{ \Lambda_H^2}  \right] \ . 
\end{eqnarray}
The effective theory features small self couplings, even though it stems from a highly nonperturbative underlying gauge theory.  
%
The glueball mass receives $1/N^2$ corrections at the fundamental theory level.  
By computing the one loop corrections to the $h$ two-point function one can check that 
the effective theory correctly reproduces the expected large $N$ corrections. Of course, the effective Lagrangian is not renormalizable in the usual sense but because the $1/N$ and $\partial^2 /\Lambda_H^2$ ordering it is possible to organise and subtract the divergences order by order in this double expansion.  
Futhermore the coefficients of this effective theory can be determined, for a given underlying Yang-Mills gauge theory, via lattice simulations \cite{Lucini:2013qja}.

\section{Large N Scalars for Dynamical Higgs Models }
\label{DESB}
We now extend the framework presented above in order to introduce consistent effective descriptions of dynamical Higgs models . We are not concerned with fitting the latest experimental data but focus instead on elucidating the associated large $N$ dynamics  and effective theory structure. 

We shall first introduce different examples and then, for each of these example we compute the  electroweak precision observables in Section~\ref{ST}, more specifically the $S$ and $T$ parameters \cite{Peskin:1990zt}.

\subsection{The Dynamical Higgs as the lightest Large N glueball }

We start by considering the logical possibility that the dynamical Higgs state is the lightest glueball state of a new fundamental composite theory. Besides pure Yang-Mills one can also consider theories with matter displaying large distance conformality and then add an explicit source of conformal breaking, such as fermion masses. This has been show to occur via lattice simulations \cite{Bursa:2011ru} and via controllable perturbative examples \cite{Grinstein:2011dq,Antipin:2011aa}. 

Within this scenario one can envision the newly discovered particle at the Large Hadron Collider (LHC) to be  the lightest glueball state of a new Yang-Mills theory with a new $N$-independent string tension proportional to $\Lambda_H$. The scale is not automatically related, here, with the electroweak symmetry breaking scale $v\simeq 246$~GeV or $4\pi v$. Therefore dynamical spontaneous breaking of the electroweak symmetry is triggered by either another strongly coupled sector or, if within the same theory, via a distinct dynamically induced chiral symmetry scale\footnote{We remind the reader that in theories with an intact centre group the confining and the chiral scale are well separated \cite{Mocsy:2003qw}.}. Since the state $h$ is a singlet with respect to the Standard Model (SM)  symmetries we can write  
\begin{eqnarray}
{\cal L}_{\rm Glueball-Higgs} &=& {\cal L}_{\overline{\rm SM}}
+\left(1+\frac{2 r_\pi}{N \Lambda_H}h+\frac{s_\pi}{N^2\Lambda_H^2}h^2\right)\frac{v^2}{4}{\rm Tr}\ D_\mu U^\dagger D^\mu U  \nonumber \\
&+&\frac{1}{2} \partial_{\mu} h \partial^{\mu} h 
 -  \frac{m^2_h}{2} h^2 \left[ 1 +  \frac{V_{0,1}}{N}\frac{h}{ \Lambda_H}   +\frac{V_{0,2}}{N^2}\frac{h^2}{ \Lambda_H^2}  \right] \
 \nonumber \\
&-&
\ m_t\left(1+\frac{r_t}{N\Lambda_H}h\right)
\Bigg[\overline{q}_L\ U\ \Bigg(\frac{1}{2}+T^3\Bigg)\ q_R + {\rm h.c.} \Bigg] \nonumber \\
&-& m_b\left(1+\frac{r_b}{N\Lambda_H}h\right)
\Bigg[\overline{q}_L\ U\ \Bigg(\frac{1}{2}-T^3\Bigg) \ q_R + {\rm h.c.} \Bigg] + \cdots \nonumber \\
&+&{\cal O}\left(\frac{1}{4\pi v},\frac{\partial^2}{\Lambda_H^2}\right)
\label{GBSM}
\end{eqnarray}
where $ {\cal L}_{\overline{\rm SM}}$ is the SM Lagrangian without Higgs and Yukawa sectors, the ellipses denote Yukawa interactions for SM fermions other than the top-bottom doublet $q\equiv(t,b)$, and ${\cal O}(1/\Lambda_H)$ includes higher-dimensional operators, which are suppressed by powers of $1/\Lambda_H$. In this Lagrangian $U$ is the usual exponential map of the Goldstone bosons produced by the breaking of the electroweak symmetry, $U=\exp\Big(i 2 \pi^a T^a/v\Big)$, with covariant derivative $D_\mu U\equiv \partial_\mu U -i g W^a_\mu T^a U + i g^\prime U B_\mu T^3$, $2T^a$ are the Pauli matrices, with $a=1,2,3$. The kinetic term and potential of the SM Higgs have been replaced by the effective theory for the lightest glueball state.  The tree-level SM is recovered for $
r_\pi =r_t=r_b= N\frac{\Lambda_H}{v}$ and $
s_\pi=  N^2\frac{\Lambda_H^2}{v^2}$. Here we will keep these couplings of order unity. Also we have not speculated on the hidden sector providing the link between the new glueball theory and the SM sector, but required  it to respect the large $N$ counting for the insertion of an extra Glueball-Higgs. We have also ordered the higher derivatives on $h$ such that they are subleading when compared to the $1/N$ operators retained here. 

If we consider the infinite number of colours limit of the new Yang-Mills theory first we arrive at a perturbative self-interacting glueball state coupled to the SM also via perturbative couplings. We have, therefore, at our disposal an organisation structure that allows to go beyond the tree-level  \cite{Hansen}. We shall investigate the dependence on the number of new colors $N$ in the section dedicated to the electroweak parameters.

 \subsection{The Large N physics of the Dynamical Higgs}

In time-honoured models of dynamical electroweak symmetry breaking \cite{Weinberg:1975gm,Susskind:1978ms}  the Higgs can be identified with a fermion-antifermion meson\footnote{This does not always have to be the case, meaning that the lightest state can be, in principle, made by multi-fermion states.}. Depending on the new  fermion representation with respect to the underlying gauge group one can have different large $N$ countings \cite{Sannino:2009za} such as the Corrigan and Ramond one \cite{Corrigan:1979xf}. The counting is incorporated directly in the pion decay constant $F_\Pi^2 = d(R) \Lambda_{TC}^2$ with $d(R)$ the dimension of the technicolor theory and $\Lambda_{TC}$ an intrinsic scale independent on the number of colours \cite{Sannino:2007yp,Kiritsis:1989ge}.    

Let's assume for definitiveness that we have an $SU(\overline{N})$ underlying theory featuring a doublet of techniquarks transforming according to the representation $R$ of the composite theory and therefore we can set $v = F_\Pi (\overline{N})$. For a generic $N$ it is sufficient to replace $v$ with $ v \,\sqrt{d/\overline{d}}$. For example for the fundamental representation $d = N$ and $\overline{d} = \overline{N}$.  
Differently from the previous Glueball-Higgs example here also the pion sector is affected by the large $N$ scaling since the self-interactions among the composite pions are also controlled by $N$. Choosing for definitivness the underlying fermions to belong to the fundamental representation we have  
\begin{eqnarray}
{\cal L}_{TC}(N)&=& {\cal L}_{\overline{\rm SM}}
+ \frac{N}{\overline{N}}\left(1+\frac{2 r_\pi}{v} \sqrt{\frac{\overline{N}}{N}} h+\frac{\overline{N}}{N}\frac{s_\pi}{v^2}h^2\right)\frac{v^2}{4}{\rm Tr}\ D_\mu U^\dagger D^\mu U  \nonumber \\
&+&\frac{1}{2} \partial_{\mu} h \partial^{\mu} h 
-  \frac{m^2_h}{2} h^2 \left[ 1 +  \frac{\sqrt{\overline{N}}{V_{0,1}}}{\sqrt{N}}\frac{h}{ v}   +\frac{\overline{N} V_{0,2}}{N}\frac{h^2}{ v^2}  \right] \
 \nonumber \\
&-&
\ \frac{y_t \, v} { \sqrt{2\overline{N}}} \sqrt{N}\left(1+\frac{\sqrt{\overline{N}}r_t}{\sqrt{N}v}h\right)
\Bigg[\overline{q}_L\ U\ \Bigg(\frac{1}{2}+T^3\Bigg)\ q_R + {\rm h.c.} \Bigg] \nonumber \\
&-& \frac{y_b \, v }{\sqrt{2\overline{ N}}} \sqrt{N} \left(1+\frac{\sqrt{\overline{N}}r_b}{\sqrt{N}v}h\right)
\Bigg[\overline{q}_L\ U\ \Bigg(\frac{1}{2}-T^3\Bigg) \ q_R + {\rm h.c.} \Bigg] + \cdots \nonumber \\
&+&{\cal O}\left(\frac{1}{4\pi v},\frac{\partial^2}{v^2}\right) \ . 
\label{eq:TC}
\end{eqnarray}
We have therefore: 
\begin{eqnarray}
m_W^2 = g^2 v^2\frac{N}{4 \overline{N}} \ , \qquad m_q = y_q v  \sqrt{\frac{N}{2\overline{N}}} \ , 
\end{eqnarray}
with $q$ a given quark,  $g$ the weak coupling and $y_q$ the Yukawa coupling. Substituting $v$ in terms of $m_W$ and the weak coupling via 
\begin{eqnarray}
\sqrt{\frac{N}{\overline{N}}} v = 2 \frac{m_W}{g} \ , 
\end{eqnarray}
in the effective Lagrangian we arrive at 
 \begin{eqnarray}
{\cal L}_{TC}(N)&=& {\cal L}_{\overline{\rm SM}}
+ \left(1+g\frac{2 r_\pi}{2 m_W}  h+ g^2\frac{s_\pi}{4 m_W^2}h^2\right)\frac{m^2_W}{g^2}{\rm Tr}\ D_\mu U^\dagger D^\mu U  \nonumber \\
&+&\frac{1}{2} \partial_{\mu} h \partial^{\mu} h 
-  \frac{m^2_h}{2} h^2 \left[ 1 +   g V_{0,1}\frac{h}{2m_W}   + g^2 V_{0,2} \frac{h^2}{ 4 m_W^2}  \right] \
 \nonumber \\
&-&
\  m_t \left(1+ g \frac{r_t}{2 m_W} h\right)
\Bigg[\overline{q}_L\ U\ \Bigg(\frac{1}{2}+T^3\Bigg)\ q_R + {\rm h.c.} \Bigg] \nonumber \\
&-& m_b\left(1+g \frac{r_b}{2 m_W} h \right)
\Bigg[\overline{q}_L\ U\ \Bigg(\frac{1}{2}-T^3\Bigg) \ q_R + {\rm h.c.} \Bigg] + \cdots \nonumber \\
\label{eq:TC}
\end{eqnarray}
Nicely the $1/\sqrt{N}$ cost of introducing an extra power of the composite field $h$ is monitored by the corresponding power in $g$.

 \subsection{The large $N$ dynamical pseudo-dilaton } 
It is possible to imagine that the Higgs state of the SM is associated to the spontaneous breaking of a conformal symmetry.  There are several possible realisations according to which the breaking of the conformal dynamics can be either associated to a nonperturbative sector  \cite{Sannino:1999qe,Goldberger:2008zz} or a perturbative one \cite{Grinstein:2011dq,Antipin:2011aa}. 

The large $N$ counting together with the request to satisfy the  conformal relations can be both ensured by imposing 
\begin{eqnarray} 
 N\Lambda_H = f \ , \qquad r_q= r_\pi = s_\pi = 1 \ ,
\end{eqnarray}
with $f$, in general, a new scale. 
The Lagrangian in \eqref{GBSM} then becomes:
\begin{eqnarray}
{\cal L}_{\rm Dilaton}(N) &=& {\cal L}_{\overline{\rm SM}}
+\left(1+\frac{2 h}{N \Lambda_H}+\frac{h^2}{N^2\Lambda_H^2}\right)\frac{v^2}{4}{\rm Tr}\ D_\mu U^\dagger D^\mu U  \nonumber \\
&+&\frac{1}{2} \partial_{\mu} h \partial^{\mu} h 
-  \frac{m^2_h}{2} h^2 \left[ 1 +  \frac{V_{0,1}}{N}\frac{h}{ \Lambda_H}   +\frac{V_{0,2}}{N^2}\frac{h^2}{ \Lambda_H^2}  \right] \
 \nonumber \\
&-&
\ m_t\left(1+\frac{h}{N\Lambda_H}\right)
\Bigg[\overline{q}_L\ U\ \Bigg(\frac{1}{2}+T^3\Bigg)\ q_R + {\rm h.c.} \Bigg] \nonumber \\
&-& m_b\left(1+\frac{h}{N\Lambda_H}\right)
\Bigg[\overline{q}_L\ U\ \Bigg(\frac{1}{2}-T^3\Bigg) \ q_R + {\rm h.c.} \Bigg] + \cdots \nonumber \\
 \label{GBSM}
\end{eqnarray}
 This framework can be immediately extended to any dilatonic-like interpretation of the state $h$, such as the one coming from a near conformal-like technicolor dynamics where $f$ ($\Lambda_H$)  is identified with $v$. Because we would like to investigate the explicit $N$ dependence we hold fix the $N$ independent scale $\Lambda_H$. Clearly this limit corresponds to the Glueball-Higgs case with extra constraints for the $h$ couplings.  

We are now ready to investigate the first consequences of the large $N$ counting.

\section{  S and T parameters}
\label{ST}
As a relevant application of the formalism introduced above we study two important correlators, i.e. the $S$ and $T$ parameters \cite{Peskin:1991sw} for the three different types of dynamical Higgs models discussed above.  

Defining by $S$ the difference between $S$ in the full theory $S_{\rm theory}$,  and the value of $S$ in the SM, i.e. $S_{\rm SM} $ we arrive at  \cite{Foadi:2012ga}  
\begin{eqnarray}
S &=& \left( 1 -  \frac{\kappa_1^2}{4}\right) \left[\frac{f(m_Z/m_h)}{6\pi}+\frac{1}{12\pi}\log\frac{(4\pi\Lambda_H)^2}{m_h^2}
+\frac{5}{72\pi}\right] + 16\pi\ c(4\pi \Lambda_H) \nonumber \\
&+& \frac{1}{12\pi}\log\frac{m_h^2}{m_{h,{\rm ref}}^2} + \frac{f(m_Z/m_{h,{\rm ref}})-f(m_Z/m_h)}{6\pi}\ .
\label{eq:S0}
\end{eqnarray}
where $\kappa_1$ is the coefficient of the linear term in $h/v$ multiplying the operator ${\rm Tr} \left[ D_{\mu} U D^{\mu} U^{\dagger}\right]$ which in the SM is equal to two, and 
\begin{equation}
f(x)\equiv\frac{2x^2+x^4-3x^6+(9x^4+x^6)\log x}{(1-x^2)^3} \ .
\end{equation}  
The $c$ term in \eqref{eq:S0} is a needed counterterm. For the $T$ parameter one obtains \cite{Foadi:2012ga}
 \begin{eqnarray}
T &=& -\frac{3}{16\pi \cos^2\theta_w}\left(1-\frac{\kappa_1^2}{4}\right)\log\frac{(4\pi \Lambda_H)^2}{m_h^2} + T_{\rm SM}^h(m_h) - T_{\rm SM}^h(m_{h,{\rm ref}}) \ ,
\label{eq:T}
\end{eqnarray}
 where we have absorbed the finite part of the counterterm in the actual value of $\Lambda_H$. $m_{h,{\rm ref}}$ is the reference value of the Higgs mass, $\theta_w$ is the Weinberg angle,  and $T_{SM}^h(m_h)$ is given in equation (22) of \cite{Foadi:2012ga}. 
 
 We can now determine the large $N$ behaviour of these relevant parameters coming from the various large $N$ dynamical Higgs models introduced earlier.

\subsection{S and T for the Large  $N$ Glueball and Dilaton Higgs Models}

In the glueball case we have 
\begin{eqnarray}
\kappa_1^{GB} = 2\frac{v r_\pi}{N\Lambda_H}  \ , 
\end{eqnarray}
 and the SM limit is recovered when $N \Lambda_H =  v r_\pi$. The precision parameters are then
\begin{eqnarray}
S_{GB} &=& \left( 1 - \frac{v^2 r_\pi^2}{N^2\Lambda_H^2}\right) \left[\frac{f(m_Z/m_h)}{6\pi}+\frac{1}{12\pi}\log\frac{(4\pi \Lambda_H)^2}{m_h^2}
+\frac{5}{72\pi}\right] + 16\pi\ c_{GB}(4\pi \Lambda_H) \nonumber \\&+& \frac{1}{12\pi}\log\frac{m_h^2}{m_{h,{\rm ref}}^2} + \frac{f(m_Z/m_{h,{\rm ref}})-f(m_Z/m_h)}{6\pi}\ ,
 \label{eq:SGB}
\end{eqnarray}
and
 \begin{eqnarray}
T_{GB} &=& -\frac{3}{16\pi \cos^2\theta_w}\left( 1 - \frac{v^2 r_\pi^2}{N^2\Lambda_H^2}\right) \log\frac{(4\pi \Lambda_H)^2}{m_h^2} + T_{\rm SM}^h(m_h) - T_{\rm SM}^h(m_{h,{\rm ref}}) \ .
\label{eq:TGB}
\end{eqnarray}
We have chosen $m_h =125$~GeV and $m_{h,{\rm ref}} = 117$~GeV and $c_{GB}(4\pi\Lambda_H)$ is a counterterm that depends on the specific underlying theory. For the Glueball-Higgs theory we absorbe the unknown counterterm in the definition of $\Lambda_H$ and therefore set it to zero in the numerical evaluation. We expect that higher glueball states will not modify the $N$ counting.  

The first observation is that if the glueball-Higgs scale $\Lambda_H$ is larger than the electroweak scale $ v$  there is a positive contribution to the $S$ parameter and an associated negative one for the $T$ parameter. Vice-versa we observe a reduction (increase) of the $S$  ($T$) parameter if $\Lambda_H$ is smaller than $v$. This is an intriguing general result given the fact that the scale of compositeness is $4\pi \Lambda_H$ can be kept above the electroweak scale.

\begin{figure}[h]
\begin{center}
\includegraphics[width=.46\textwidth]{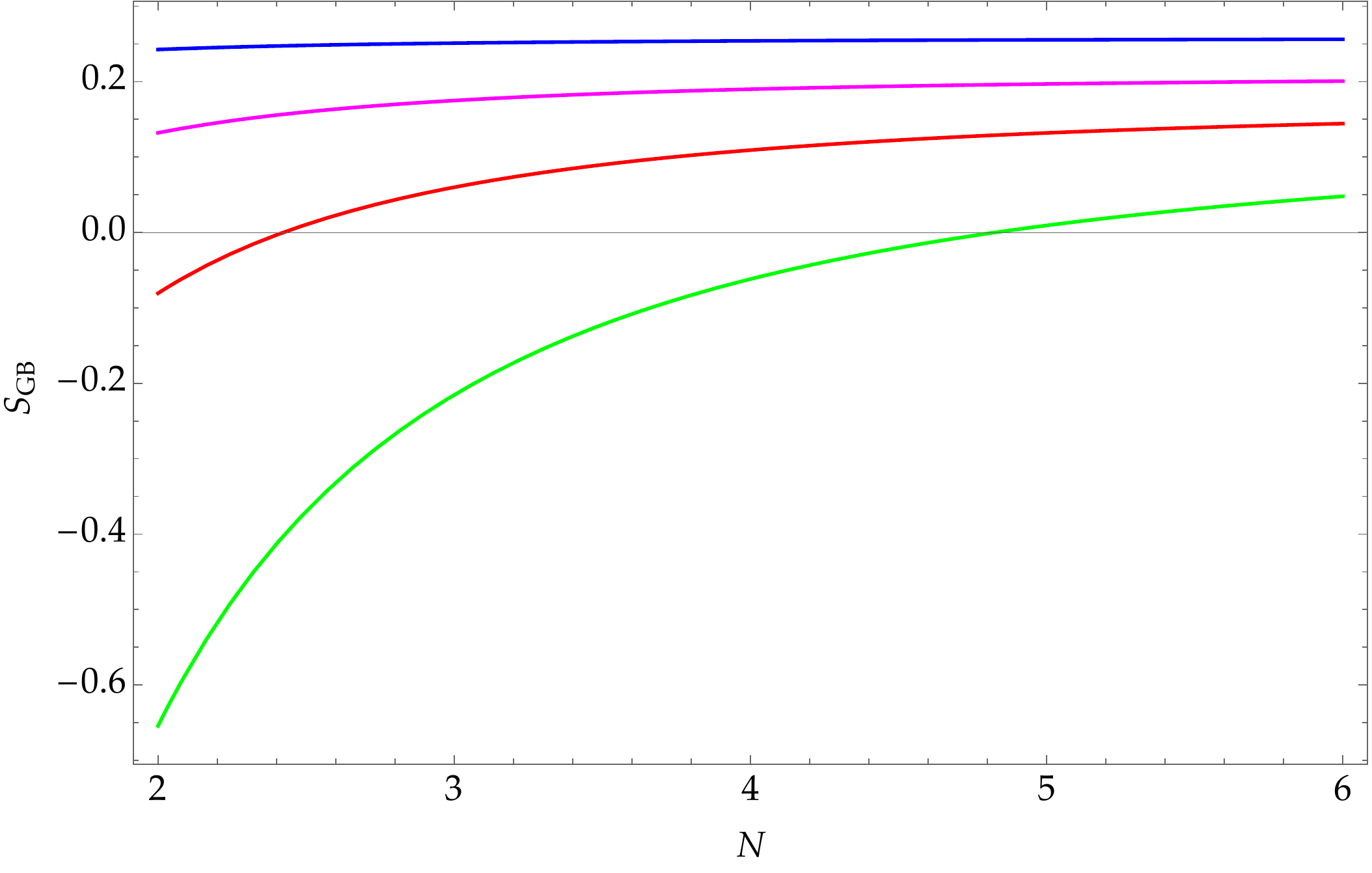} \hskip .4cm
\includegraphics[width=.46\textwidth]{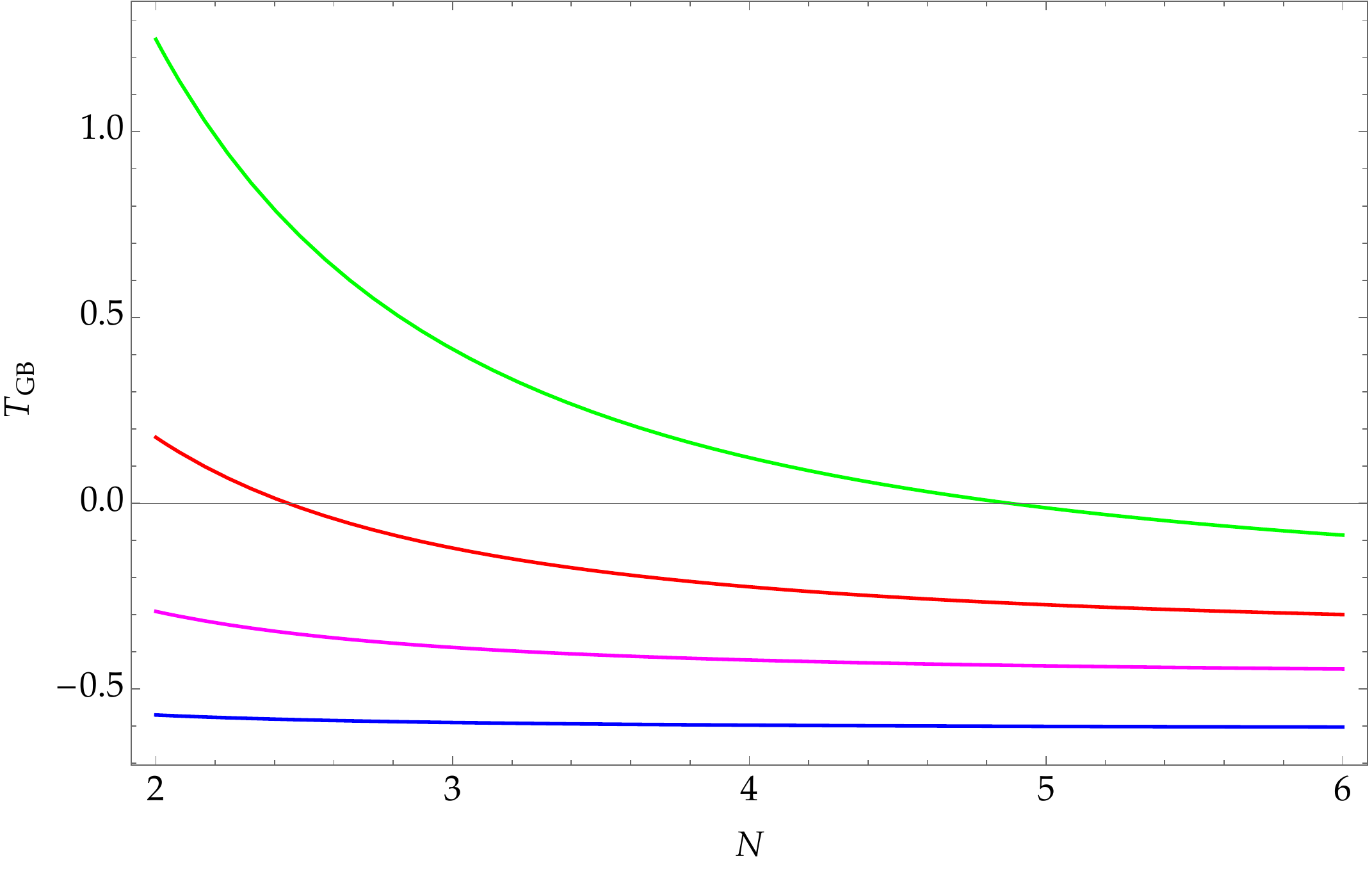} 
\caption{We show the dependence on the number of underlying glueball-Higgs colors for the $S$ (left-panel) and $T$(right-panel)   for $\Lambda_H = 500$ GeV (blue curve), $200$ GeV (magenta),  $100$ GeV (red) and $50$ GeV (green). The composite scale $4\pi \Lambda_H$ is always higher than the electroweak scale of $246$ GeV, and further assumed $r_\pi \approx 1$.  }
\label{SandTGB}
\end{center}
\end{figure}

Increasing $N$ while keeping fixed $\Lambda_H$ and $r_\pi$ one arrives at the following $N$-independent results  
 \begin{eqnarray}
\lim_{N\rightarrow \infty}S_{GB} &=&  \left[\frac{f(m_Z/m_h)}{6\pi}+\frac{1}{12\pi}\log\frac{(4\pi \Lambda_H)^2}{m_h^2}
+\frac{5}{72\pi}\right] + 16\pi\ c_{GB}(4\pi \Lambda_H) \nonumber \\&+& \frac{1}{12\pi}\log\frac{m_h^2}{m_{h,{\rm ref}}^2} + \frac{f(m_Z/m_{h,{\rm ref}})-f(m_Z/m_h)}{6\pi}\ ,
 \label{eq:SGB-LargeN}
\end{eqnarray}
and 
 \begin{eqnarray}
\lim_{N\rightarrow \infty}T_{GB} &=& -\frac{3}{16\pi \cos^2\theta_w} \log\frac{(4\pi \Lambda_H)^2}{m_h^2} + T_{\rm SM}^h(m_h) - T_{\rm SM}^h(m_{h,{\rm ref}}) \ .
\label{eq:TGB-LargeN}
\end{eqnarray} 
The corrections appear at ${\cal O}(N^2)$. 

\begin{figure}[h]
\begin{center}
\includegraphics[width=.5\textwidth]{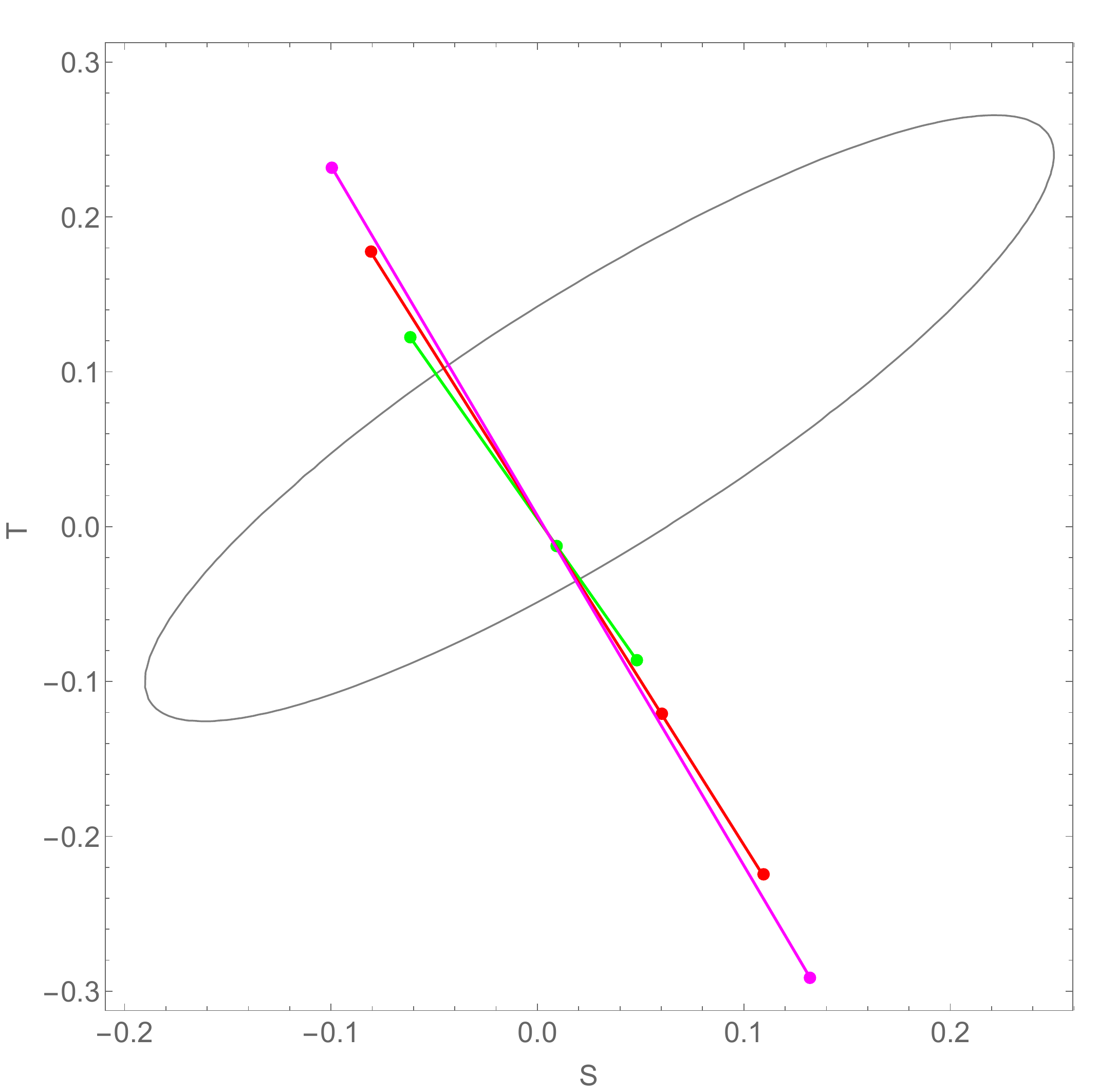} 
\caption{Comparison with the precision electroweak constraints for the glueball-Higgs for $\Lambda_H = 200$ GeV (magenta) and $N=1,2$,  $100$ GeV (red) and $N=2,3,4$, and $50$ GeV (green) for $N=4,5,6$. The composite scale $4\pi \Lambda_H$ is always higher than the electroweak scale of $246$ GeV.  The left most point on each curve corresponds to the smallest $N$, and further assumed $r_\pi = 1$.  }
\label{STvsGB}
\end{center}
\end{figure}

In Figure \ref{SandTGB} we plot $S$ and $T$ as function of the number of colors for different values of $\Lambda_H$. Because the corrections are in $1/N^2$ the large $N$ limit is approached quickly. We have also assumed $r_\pi \approx 1$ that is its natural order of magnitude and, in any event, can be partially reabsorbed in $\Lambda_H$. We compare the result with the experimental value of precision data in Fig.~\ref{STvsGB} for  $\Lambda_H = 200$ GeV (magenta) and $N=1,2$. In red we have   $\Lambda_H = 100$ GeV  and $N=2,3,4$. Finally we plot the $50$ GeV (green) case for $N=4,5,6$. The composite scale $4\pi \Lambda_H$ is always higher than the electroweak scale of $246$ GeV.  The left most point on each curve corresponds to the smallest $N$. The experiments prefer smaller values of $\Lambda_H$ with $N$ in the range two to four.  Larger values of $\Lambda_H$ require $N$ to be away from the large N limit and therefore we cannot conclude on the viability of the $\Lambda_H = 200$ GeV case. Increasing further $\Lambda_H$ it is clearly not preferred by  precision observables. If, therefore, a Glueball-Higgs model does describe the Higgs we expect soon new states to be discovered with masses in the range $600 - 1200$ GeV. 

We stress that by requiring to be in agreement with precision measurements the couplings of the Higgs to the standard model gauge bosons are also close to the experimental values. This occurs because the product $N \Lambda_H$ is constrained to be near the electroweak scale. 

 For the Dilaton-Higgs example we have
\begin{eqnarray}
\kappa_1^{\rm Dilaton} = 2 \frac{v }{N\Lambda_H}  \ ,
\end{eqnarray}
 that corresponds to the results above but now with $r_\pi $ exactly equal to one.

\subsection{S and T for the large $N$ Dynamical Higgs} 
\begin{figure}[h!]
\begin{center}
\includegraphics[width=.33\textwidth]{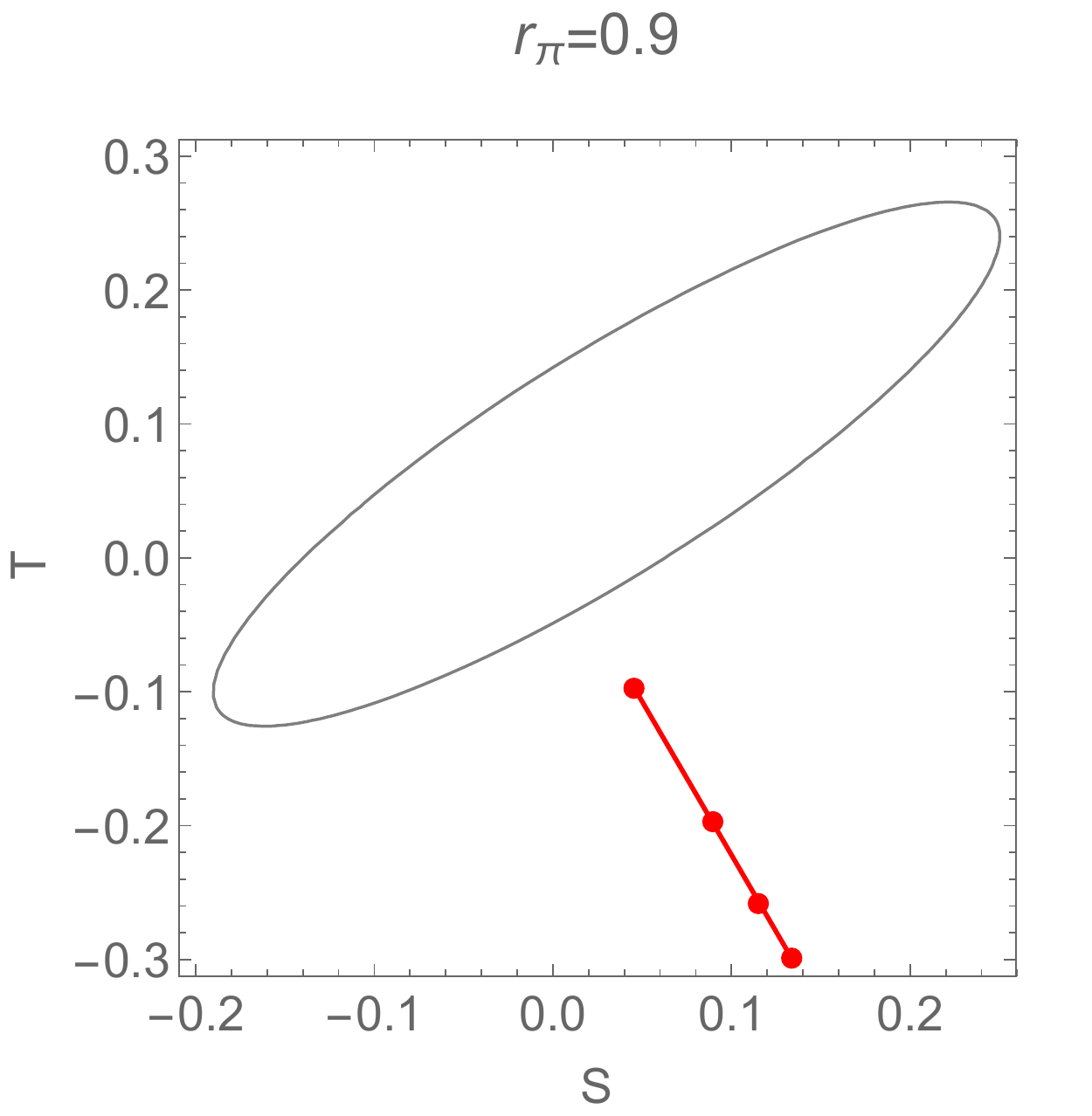} 
\includegraphics[width=.33\textwidth]{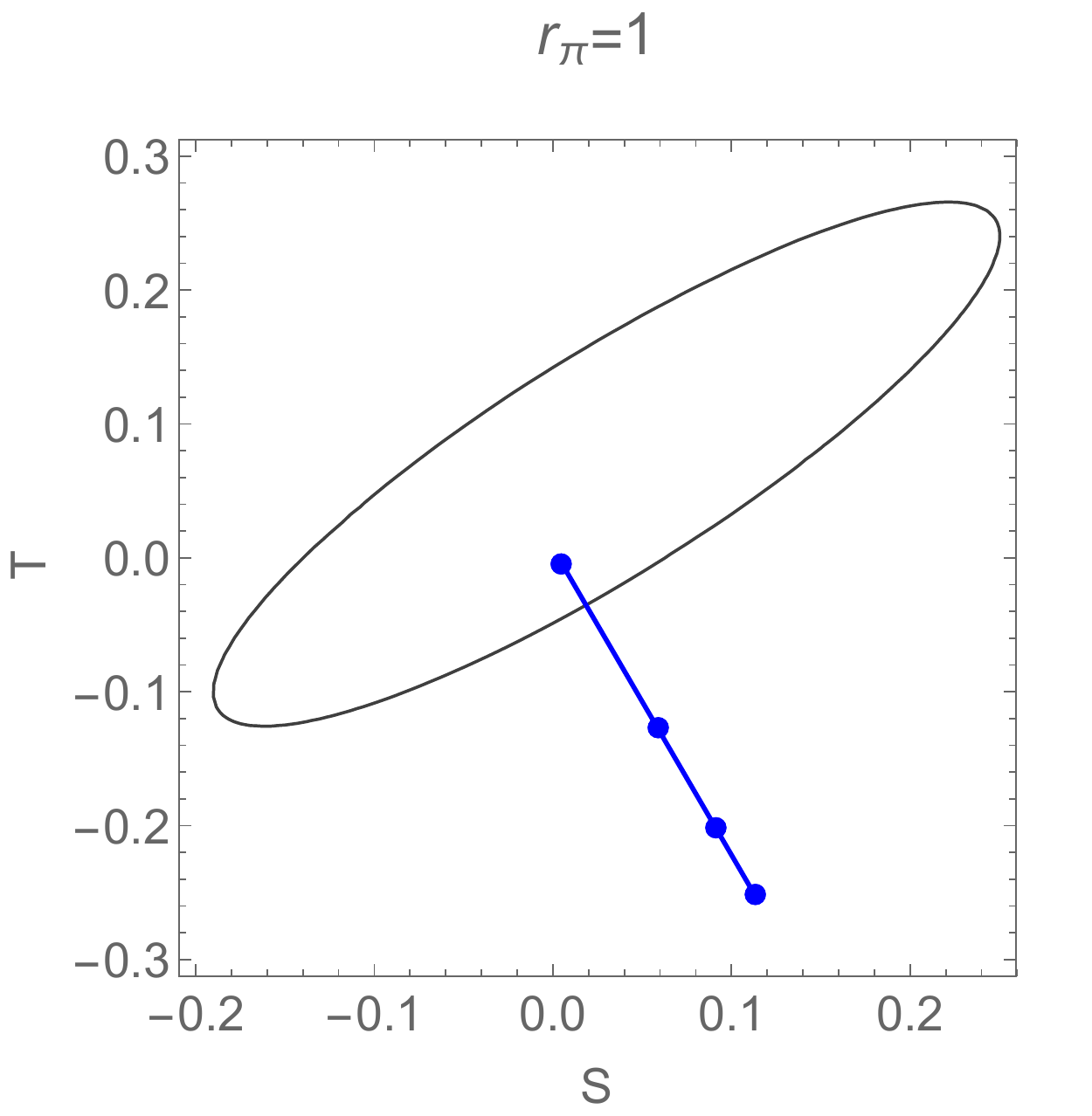} 
\includegraphics[width=.33\textwidth]{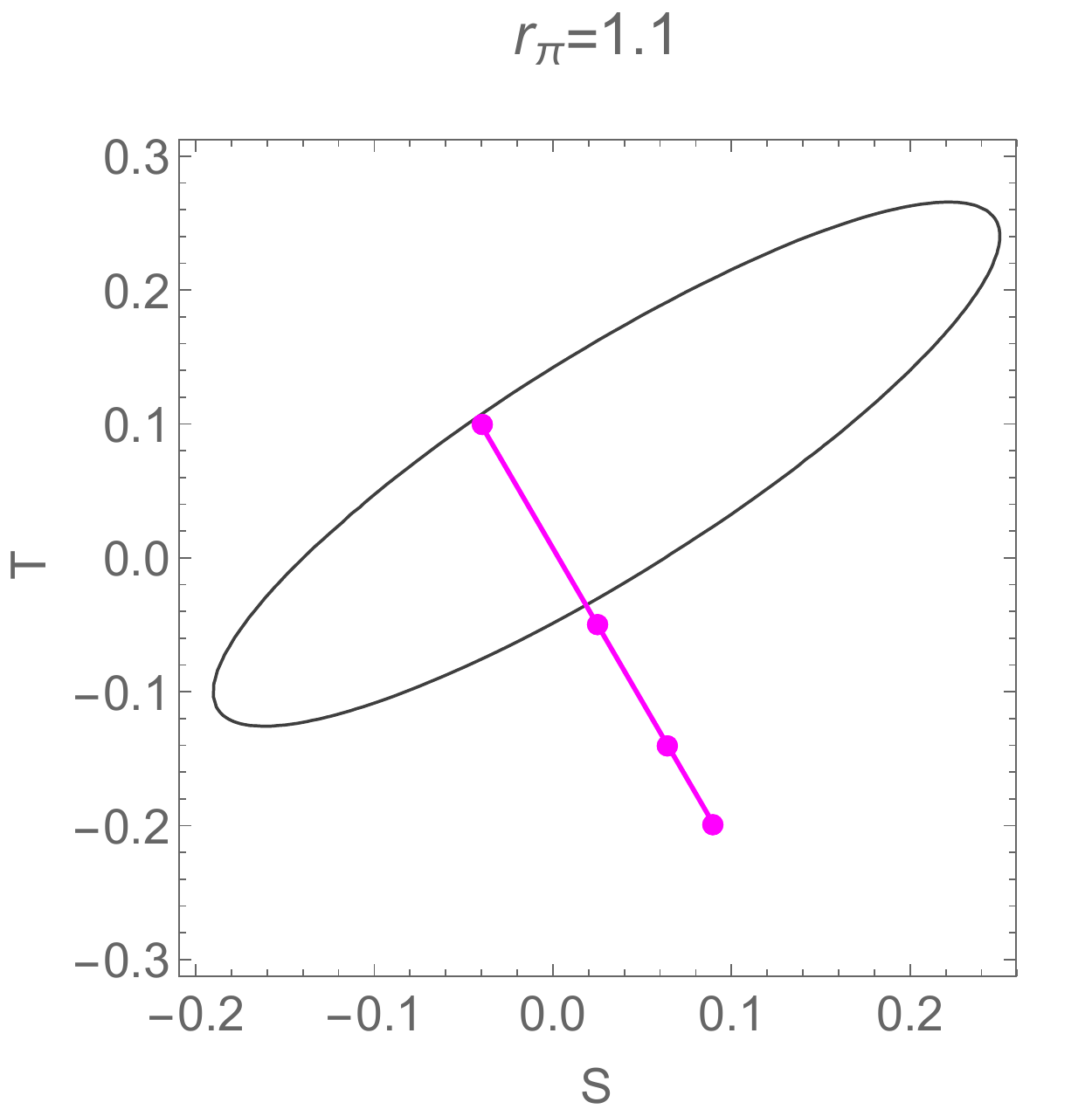} 
\caption{Comparison with the precision electroweak constraints for the dynamical-Higgs for $ v = 246$ GeV and $N=3,4,5,6$. The left most point on each curve corresponds to the smallest $N$, and further assumed $r_\pi = 0.9$ (left panel),  $r_\pi =1 $ (central panel), $r_\pi \approx 1.1$ (right panel). }
\label{STvsDH}
\end{center}
\end{figure}

It is interesting to explore what happens for the large $N$ dynamical Higgs. The main difference with respect to the previous case is that the electroweak scale and the dynamical Higgs scale are now identified. Among the possible underlying models that can lead to this kind of effective dynamics there are time-honoured examples such as minimal models of (near-conformal) technicolor \cite{Appelquist:1999dq,Sannino:2004qp,Dietrich:2005jn}. 
\begin{eqnarray}
\kappa_1^{TC} =  2  r_\pi \sqrt{\frac{ {\overline{N}}}{N}} \ ,
\end{eqnarray}
yielding
 \begin{eqnarray}
S_{TC} &=& \left( 1 - \frac{\overline{N}}{N} r_\pi^2\right) \left[\frac{f(m_Z/m_h)}{6\pi}+\frac{1}{12\pi}\log\frac{( 4\pi v)^2}{m_h^2} 
+\frac{5}{72\pi}\right] + 16\pi\ c(4\pi v)  \\&+& \frac{1}{12\pi}\log\frac{m_h^2}{m_{h,{\rm ref}}^2} + \frac{f(m_Z/m_{h,{\rm ref}})-f(m_Z/m_h)}{6\pi}\ ,
\label{eq:STC}
\end{eqnarray}
and
 \begin{eqnarray}
T_{TC} &=& -\frac{3}{16\pi \cos^2\theta_w} \left( 1 - \frac{\overline{N}}{N} r_\pi^2\right)\log\frac{(4\pi v)^2}{m_h^2} + T_{\rm SM}^h(m_h) - T_{\rm SM}^h(m_{h,{\rm ref}}) \ .
\label{eq:TTC}
\end{eqnarray}
Differently from the Glueball-Higgs case we have at our disposal only the $N$ dependence of the effective coupling which goes as $1/N$ for the fundamental representation (chosen here) or as 
$1/{N(N\pm 1)}$ for two-index representations \cite{Sannino:2007yp}.  We show in Fig.~\ref{STvsDH} the comparison with the precision electroweak constraints for the dynamical-Higgs for $ v = 246$ GeV and $N=3,4,5,6$. The left most point on each curve corresponds to the smallest $N$, and further assumed $r_\pi = 0.9$ (left panel),  $r_\pi =1 $ (central panel), $r_\pi \approx 1.1$ (right panel). It is clear from the results that it is possible to abide the electroweak precision constraints for larger number of colours provided that $r_\pi$ is larger than in the SM.

The computation elucidate another important point, i.e. that at large $N$ the leading corrections, expected to be proportional to $N$, do not come from the dynamical-Higgs sector but rather from the tree-level exchange of spin one resonances \cite{Foadi:2012ga}. A simple way to understand this point is to observe that the dynamical-Higgs corrections appear, at the effective Lagrangian level, from loops of $h$ and technipions, i.e. the longitudinal components of the SM gauge bosons. At the more fundamental level these corrections are subleading in $N$ and therefore respect the counting provided by our effective approach. Furthermore these dynamical-Higgs corrections were not taken into account in \cite{Peskin:1991sw}, have been amended in \cite{Foadi:2012ga} and here we provide the  intrinsic $N$ dependence.

  \section{Conclusions and Top Corrections}
  \label{conclusions}
We introduced effective field theories and associated counting schemes to consistently describe the lightest massive large $N$ stable composite scalar state emerging in any theory of composite dynamics.  The framework allows for systematic investigations of  composite dynamics featuring non-Goldstone (and Goldstone) scalars. As time-honoured examples  we discussed the lightest glueball state stemming from Yang-Mills theories. We further applied our effective approach to models of (near-conformal) dynamical electroweak symmetry breaking.  In particular we considered the following three possibilities: the Higgs is the lightest glueball of a new composite theory; it is a large $N$ scalar meson in models of dynamical Higgs such as technicolor, and finally we considered it to be a large $N$ pseudodilaton in the form of a conformal compensator. For each of these models, we provided the leading $N$ corrections to the precision parameters. 

We observe that it is straightforward to show that in this framework the top corrections to the Glueball and dynamical Higgs mass can be reliably estimated in the large $N$ limit by rescaling $r_t$ in  equation (4) of \cite{Foadi:2012bb,Cacciapaglia:2014uja} by the opportune power of $N$, for each model, and simultaneously replacing the cutoff scale by either $4\pi \Lambda_H$ or $ 4\pi v$. 

The results provide useful insight stemming from the large $N$ dynamics of these models and can be viewed as the stepping stone for consistent determination of quantum corrections at the effective Lagrangian level containing massive scalar states. The effective approach is directly applicable also to models of composite Goldstone Higgs dynamics \cite{Kaplan:1983fs,Kaplan:1983sm} when including the first massive scalar state \cite{Cacciapaglia:2014uja,Arbey:2015exa,Cacciapaglia:2015eqa}, as well as to investigate interesting flavour properties \cite{Ghosh:2015gpa,Altmannshofer:2015esa}.  Finally holographic studies of the spectrum and large N properties of strongly coupled theories \cite{Arean:2013tja,Alho:2013hsa,Alho:2015zua} can benefit from a model independent large N computation that can be performed with the effective theories constructed here. 

\acknowledgments
CP$^3$-Origins is partially supported by the Danish National Research Foundation grant DNRF:90.

\end{document}